\documentclass[prd,twocolumn,preprint numbers,nofootinbib, aps,showpacs]{revtex4-1}
\usepackage[utf8]{inputenc}
\usepackage[english]{babel}
\usepackage[reset, a4paper, top=3cm,bottom=3cm,left=2.5cm,right=2.5cm]{geometry}

\usepackage{subcaption}
\usepackage{graphicx, caption}
\usepackage{multirow}
\usepackage{colortbl}
\usepackage{hhline}
\usepackage{float}
\usepackage{chngcntr}
\usepackage{mathtools}
\usepackage{amssymb}
\usepackage{bm}
\usepackage{esvect}
\usepackage{commath}
\usepackage{braket}
\usepackage{hyperref}
\usepackage{siunitx}
\usepackage[symbol]{footmisc}

\usepackage{epsfig}
\usepackage{amsmath}
\usepackage{amsfonts}
\usepackage{url}
\usepackage{hhline}
\usepackage{color}
\usepackage{bm}
\usepackage{cancel}
\usepackage{times}

\begin{document}

\title{The ENUBET monitored neutrino beam and its implementation at CERN \\ \normalsize{Contribution to the 25th International Workshop on Neutrinos from Accelerators}}

\author{Leon Halić$^{1}$ - for the ENUBET collaboration*}

\affiliation{$^1$\, Center of Excellence for Advanced Materials and Sensing Devices, Ruder Bo\v{s}kovi\'c Institute, 10000 Zagreb, Croatia} 
\begin{abstract}

The ENUBET project recently concluded the R\&D for a site independent design of a monitored neutrino beam for high precision cross section measurements, in which the neutrino flux is inferred from the measurement of charged leptons in an instrumented decay tunnel. In this phase three fundamental results were obtained and will be discussed here: 1) a beamline not requiring a horn and relying on static focusing elements allows to perform a $\nu_e$ cross section measurement in the DUNE energy range with 1 \% statistical uncertainty employing $10^{20}$ $400$ GeV protons on target (pot) and a moderate mass neutrino detector of the size of ProtoDUNE; 2) the instrumentation of the decay tunnel, based on a cost effective sampling calorimeter solution, has been tested with a large scale prototype achieving the performance required to identify positrons and muons from kaon decays with high signal-to-noise ratio; 3) the systematics budget on the neutrino flux is constrained at the 1\% level by fitting the charged leptons observables measured in the decay tunnel. Based on these successful results ENUBET is now pursuing a study for a site dependent implementation at CERN in the framework of Physics Beyond Colliders. In this context a new beamline, able to enrich the neutrino flux at the energy of HK and to reduce by more than a factor 3 the needed pot, has been designed and is being optimized. The civil engineering and radioprotection studies for the siting of ENUBET in the North Area towards the two ProtoDUNEs are also in the scope of this work, with the goal of proposing a neutrino cross section experiment in 2026. The combined use of both the neutrino detectors and of the improved beamline would allow to perform cross section measurements with unprecedented precision in about 5 years with a proton request compatible with the needs of other users after CERN Long Shutdown 3. An update on the status of these studies and future plans will be presented.
 
\end{abstract}

\maketitle

\footnotetext[1]{F. Acerbi, I. Angelis, L. Bomben, M. Bonesini, F. Bramati, A. Branca, C. Brizzolari, G. Brunetti, M. Calviani, S. Capelli , M. Capitani, S. Carturan, M.G. Catanesi, S. Cecchini, N. Charitonidis, F. Cindolo, G. Cogo, G. Collazuol, F. Dal Corso, C. Delogu, G. De Rosa, A. Falcone, B.Goddard, A. Gola, D. Guffanti, L. Halić , F. Iacob, M.A. Jebramcik, C. Jollet, V. Kain, A. Kallitsopoulou, B. Kliček , Y. Kudenko, Ch. Lampoudis, M. Laveder, P. Legou , A. Longhin, L. Ludovici, E. Lutsenko, L. Magaletti, G. Mandrioli, S. Marangoni, A. Margotti, V. Mascagna, N. Mauri, J. McElwee, L. Meazza, A. Meregaglia, M. Mezzetto, M. Nessi, A. Paoloni, M. Pari, T. Papaevangelou, E.G. Parozzi, L. Pasqualini, G. Paternoster, L. Patrizii, M. Pozzato, M. Prest, F. Pupilli, E. Radicioni, A.C. Ruggeri, G. Saibene, D. Sampsonidis, A. Scanu, C. Scian, G. Sirri, R. Speziali, M. Stipčević , M. Tenti, F. Terranova, M. Torti, S.E. Tzamarias, E. Vallazza, F. Velotti, L. Votano}

\section{Introduction}

With the recent advancements in beam power and detector masses, future neutrino oscillation experiments (DUNE \cite{DUNE:2020ypp}, Hyper-Kamiokande \cite{Hyper-Kamiokande:2018ofw}, ESSnuSB \cite{Alekou:2022emd}) are no longer limited by the statistical, but by the systematic uncertainty. This systematic uncertainty mainly comes in the form of electron neutrino (the main signal in most neutrino oscillation experiments) cross section uncertainty at the GeV scale (the scale of most next generation neutrino experiments) which is currently known at $\mathcal{O}$(10-30\,\%) \cite{Branca:2021vis}. Measuring this cross section more precisely in a conventional neutrino cross section experiment is difficult, mainly due to the poor knowledge of the initial neutrino flux.

To combat this problem, ENUBET \cite{Longhin:2014yta} proposes a monitored neutrino facility which aims to reduce the uncertainty on the initial neutrino flux down to 1\,\%. This design is achieved using a conventional beamline by instrumenting the walls of the decay tunnel to observe large-angle positrons from $K_{e3}$ decays and thus constraining the $\nu_e$ flux (this is the main source of the $\nu_e$ if muon decays are suppressed by the decay tunnel length). The ENUBET concept was expanded within the context of CERN Neutrino Platform experiment NP06 \cite{Acerbi:2759849} to also constrain the $\nu_{\mu}$ flux by observing both the large-angle $\mu$ from $K_{\mu\nu}$ decays and small-angle $\mu$ from $\pi_{\mu\nu}$ decays. The former muons are observed in the decay tunnel walls similar to positrons, while the latter are observed in the instrumented hadron dump which is located right after the decay tunnel end. The main challenges for ENUBET are to design an efficient meson transfer line able to provide a clean and well collimated beam and to have a cost-effective solution for the decay tunnel instrumentation.

\section{ENUBET's meson transfer line}

\begin{figure}[ht!]
    \centering
    \includegraphics[width = \linewidth]{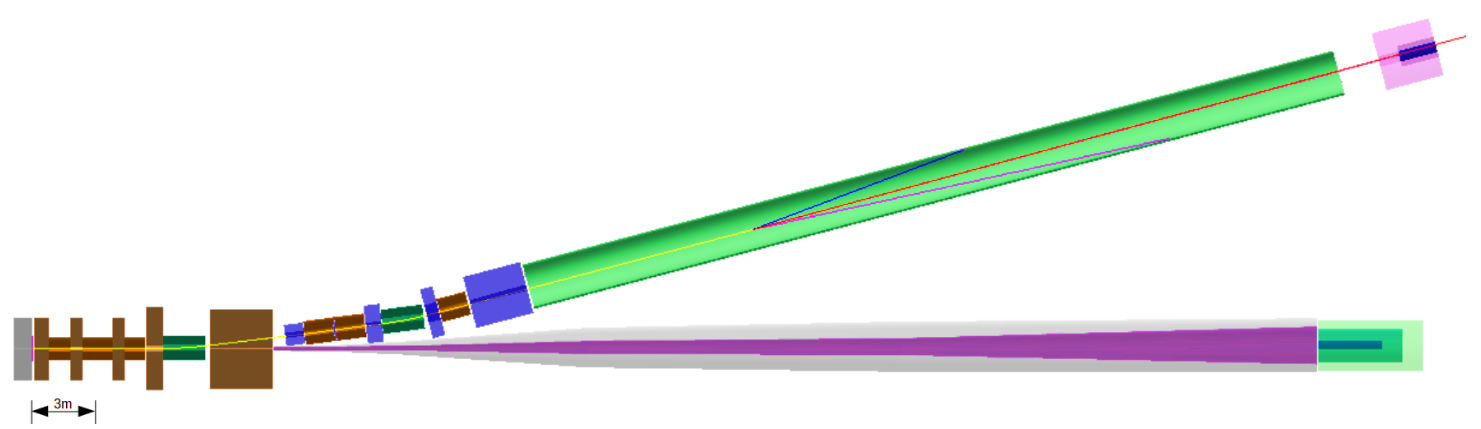}
    \caption{The final design of the ENUBET beamline. Quadrupoles are shown in orange, while bending dipoles are depicted in green. Copper collimators are reported in brown, while the elements in violet are Inermet-180 collimators. The 40 m decay tunnel is also shown, together with the pipe for non interacting protons and the proton and hadron dumps.}
    \label{beamline}
\end{figure}

The current design of the ENUBET beamline \cite{ENUBET:2023hgu}, shown in Figure \ref{beamline}, provides a narrow-band neutrino beam. It achieves this by focusing $8.5$ GeV/c positively charged mesons with a 5-10\,\% momentum bite. Charged mesons are produced from the interactions of $400$ GeV/c protons on a $70$ cm long and $3$ cm in radius graphite target. The dimensions of the target have been optimized with FLUKA to maximize the kaon yield. Similarly, the total length of the transfer line is optimized to maximize the kaon survival rate before they reach the decay tunnel. The focusing system of the secondary mesons is a fully static design with a quadrupole triplet in front of the proton target. The advantage of the static focusing system, compared to the horn-based design, is the possibility of implementing a slow proton extraction system which lasts several seconds. This static focusing and slow extraction scheme was a major breakthrough for the feasibility of ENUBET since it allows to lower the intensity of secondary mesons in the decay tunnel (thus reducing the pile-up on the instrumentation) while at the same time not losing a significant amount of statistics.

Charge and momentum selection of secondary mesons is performed by two dipoles which bend the meson beam by $14.8^{\circ}$ with respect to the incident proton beam.  Additionally, this bending significantly lowers the $\nu_e$ background at the neutrino detector from early kaon decays and from the target region.

\section{Neutrino beam}

\begin{figure}[ht!]
    \centering
    \includegraphics[width = \linewidth]{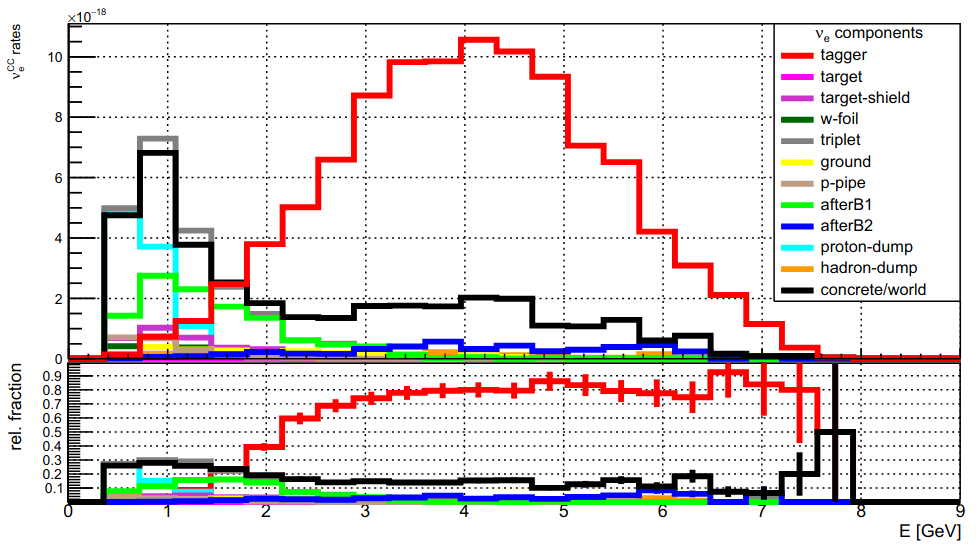}
    \caption{Energy spectrum of $\nu_e^{CC}$ interactions, with a breakdown of the neutrino components according to their production point within the ENUBET beamline. The bottom plot reports the fraction of each spectrum
relative to the total sample.}
    \label{spectrum}
\end{figure}

\begin{figure}[ht!]
    \centering
    \includegraphics[width = \linewidth]{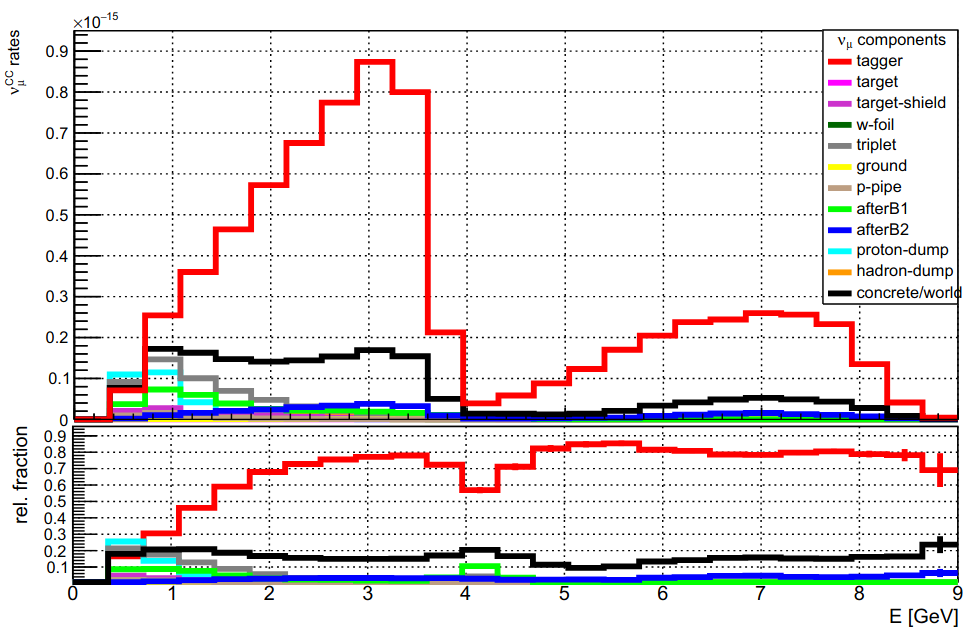}
    \caption{Energy spectrum of $\nu_{\mu}^{CC}$ interactions, with a breakdown of the neutrino components according to their production point within the ENUBET beamline. The bottom plot reports the fraction of each spectrum
relative to the total sample.}
    \label{spectrum_mu}
\end{figure}

Such a design can provide $10^4$ $\nu_e^{CC}$ and $6\times10^5$ $\nu_{\mu}^{CC}$ interactions assuming a 500-ton ProtoDUNE-like neutrino detector \cite{DUNE:2017pqt} at a $50$ m baseline and $4.5\times 10^{19}$ protons on target. The $\nu_e^{CC}$ spectrum is shown in Fig.~\ref{spectrum} where the neutrinos produced from $K_{e3}$ decays in the decay tunnel (red line) make up $68\,\%$ of all interactions in the neutrino detector above $1.5$ GeV, with additional contributions coming from the beamline and the hadron dump. In Fig. \ref{spectrum_mu}, the spectrum of $\nu_{\mu}^{CC}$ interactions is shown. We can see the typical double peak structure of narrow-band neutrino beams, where the peak below 4 GeV is dominated by $\pi$ decays, while the peak above 4 GeV is due to $K$ decays.

\begin{figure}[ht!]
    \centering
    \includegraphics[width = \linewidth]{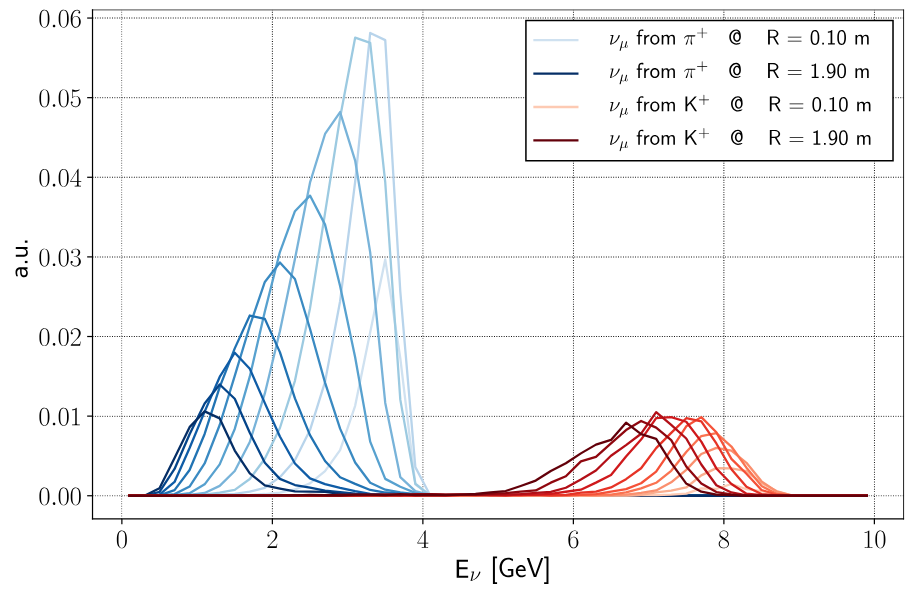}
    \caption{Energy distribution in the neutrino detector of $\nu_{\mu}$ events that originated from the decay tunnel. Each distribution shows events from a single radial interval of $\Delta r = 20$ cm.  }
    \label{nbot}
\end{figure}

Such a narrow momentum bandwidth, in conjunction with a short baseline, allows for an a-priori determination of $\nu_{\mu}$ energy using the so-called narrow-band off-axis technique. This technique exploits the correlation between the neutrino energy and the radial distance of the interaction vertex from the beam axis at the neutrino detector. This correlation is only exploitable for decays such as $\pi_{\mu\nu}$ and $K_{\mu\nu}$ due to their 2-body decays kinematics. Selecting the interactions based on radial intervals allows to determine the incoming neutrino energy with a precision given by the width of the pion peaks, shown in Figure \ref{nbot}, which ranges from $10-25\,\%$ in the DUNE energy domain.

\section{The instrumented decay tunnel and its Demonstrator}

\begin{figure}[h]
    \centering
    \includegraphics[width = \linewidth]{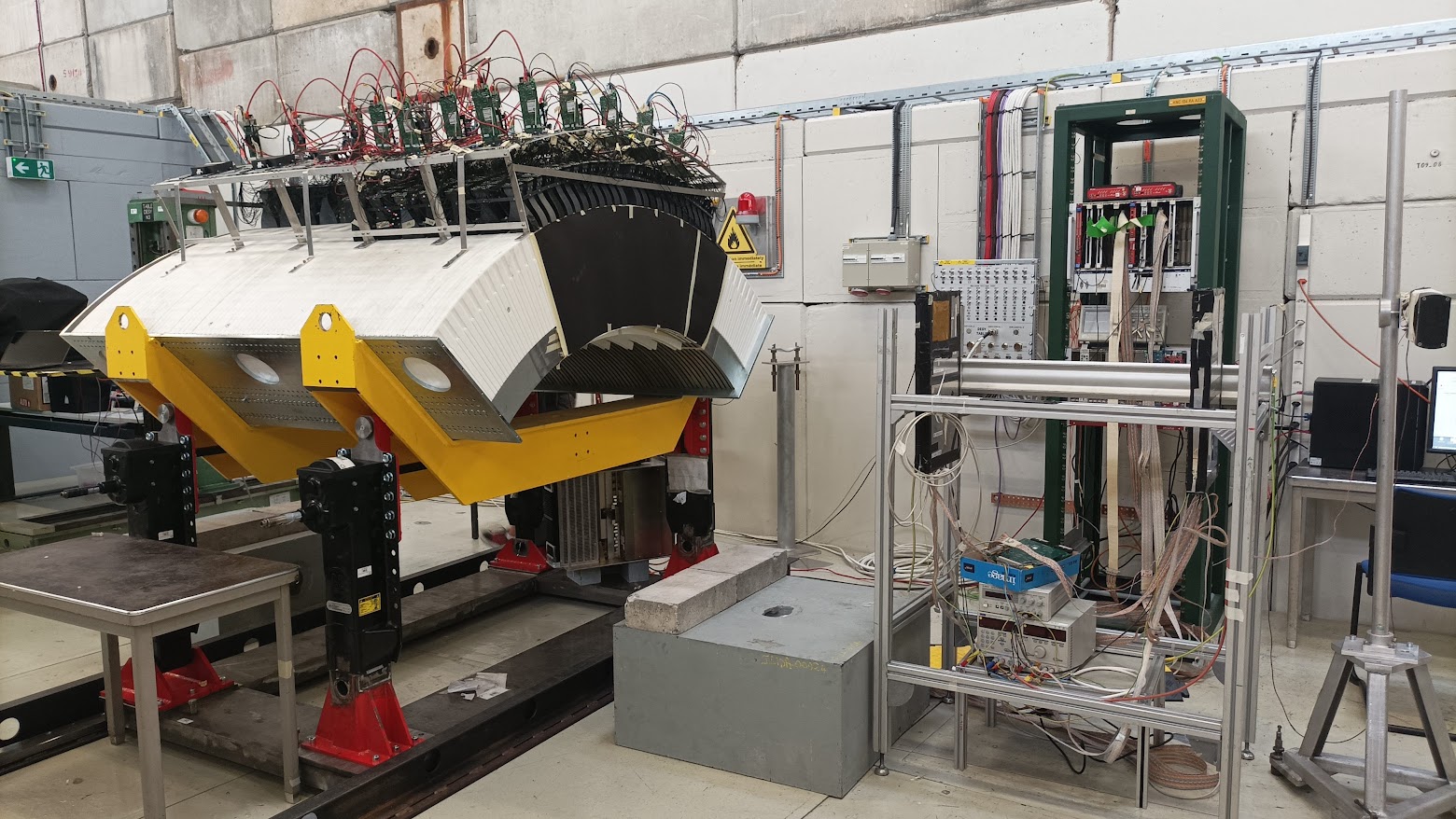}
    \caption{The ENUBET demonstrator and the testbeam setup for the 2023 testbeam at CERN-PS}
    \label{demonstrator}
\end{figure}

The instrumentation of the decay tunnel is based on a sampling calorimeter to be used for $e$/$\mu$/$\pi$ separation. The segmentation is in longitudinal (beam direction), radial and azimuthal coordinates. Each calorimetric module is a stack in the longitudinal direction of five, 0.7 cm thick, plastic scintillator tiles ($3\times3$ cm$^2$) interleaved with $1.5$ cm thick iron plates. The calorimeter is accompanied by an additional inner radial layer of scintillator tiles acting as a veto for $\gamma$/$\pi^0$ events. To test the aforementioned design, as part of the ENUBET project, a Demonstrator \cite{Acerbi:2020nwd} was constructed ($1.65$ m length, $3.5$ t mass), shown in Figure \ref{demonstrator}. The Demonstrator was partially instrumented, with currently a total of 1275 active channels and was exposed to the T9 particle beam at the CERN-PS in 2022, 2023 and 2024 \cite{Acerbi:2805716}. Full data analysis is still in progress, but the preliminary results show an energy linearity and resolution that is appropriate for $e$/$\pi$ separation in $1-3$ GeV range.

\section{Particle identification and flux systematics}

\begin{figure}[h]
    \centering
    \includegraphics[width = \linewidth]{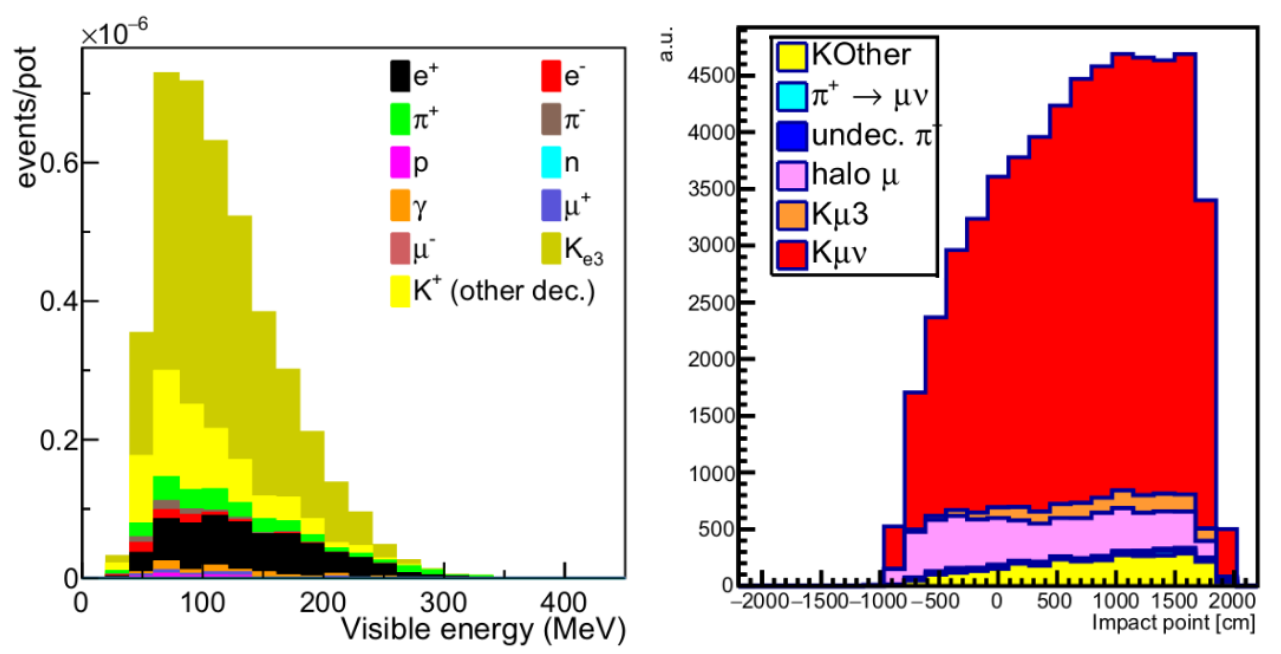}
    \caption{Distribution of observables for selected events. Left: visible energy of positrons from $K_{e3}$ signal and of background events. Right: impact point along the calorimeter of muons from $K_{\mu2}$ and $K_{\mu3}$ and of background events.}
    \label{systematics}
\end{figure}

 The full instrumentation of the decay tunnel has been simulated with GEANT4 and has been validated by data from test experiments performed at CERN. Events are identified by clustering energy deposits in space and time, while a Neural Network is used to discriminate between signal and background events by exploiting patterns in energy deposition \ref{systematics}. Using these algorithms, $22\,\%$ of positrons from $K_{e3}$ decays are reconstructed with a signal-to-noise ratio (S/N) of $\sim2$, while $34\,\%$ of muons from $K_{\mu\nu}$ decays are reconstructed with a S/N of $\sim 6$. A signal+background model is constructed for the monitored charged leptons to constrain the neutrino flux. The hadroproduction systematics, the dominant ones in the total neutrino flux uncertainty, have been included in this model as nuisance parameters and are derived from the data of NA56/SPY \cite{Arsenescu:1999zs} and NA20 \cite{Atherton:133786} experiments that used similar proton energies. The model is used to generate and fit a set of toy MC experiments from which the values for hadroproduction parameters are determined. The neutrino flux is obtained by reweighing the MC simulation by the new hadroproduction parameters. Using only the original hadroproduction data, the uncertainty on the neutrino flux is around $6\,\%$. If we include the constraints from the monitoring of the charged leptons, this uncertainty drops to $1\,\%$ which was the original ENUBET goal.

\section{The NP06/ENUBET implementation at CERN SPS}

Given the successful R\&D of the original ENUBET project and its successor NP06, there is an ongoing study in the framework of the Physics Beyond Colliders for a possible implementation of the NP06/ENUBET design at CERN. The goal is to run this implementation at the same time as Hyper-Kamiokande and DUNE to provide neutrino cross section measurements at 1\,\% uncertainty level for the GeV neutrino energies. This proposal, called "\textit{Short-Baseline neutrinos @ Physics Beyond Collider}" (SBN@PBC) is currently under study by CERN, ENUBET, NuTAG and CERN Neutrino Platform \cite{Terranova:2896594}. SBN@PBC also aims to resolve some shortcomings of the original NP06/ENUBET design, such as that it is optimized for DUNE energy range while we would also like to cover the Hyper-K energy range. An additional issue is that the original design requires more POT than SPS can deliver considering that there will be other experiments using the SPS, so lowering the required POT without significantly losing performance is also one of the SBN@PBC goals. Preliminary results of SBN@PBC show that both of these problems are solvable and that the implementation at CERN SPS is viable. Current results achieve similar performance with only 33\,\% of the POT in the original design while also being able to run at lower secondary particle momenta ($4-8.5$ GeV/c). Another possibility studied within this proposal is the time tagging of neutrinos, charged leptons and parent mesons as proposed by NuTAG \cite{Baratto-Roldan:2024bxk} to achieve superior neutrino energy resolution.

To lower the project costs, the proposal aims to use as much of the existing infrastructure as possible, this includes both the beamline magnetic elements and the neutrino detector. At this moment, there are two possible options for the neutrino detector: ProtoDUNE and WCTE \cite{Garode:2022sej} detectors. Main differences being the target material and detector technology, the former being based on liquid argon with TPC as the detector technology, while the latter is a water Cherenkov detector using standard PMTs for light collection.

In addition to the engineering and technical aspects of the implementation at CERN, a careful assessment of the physics performance and an in-depth knowledge of the assets and limitations for the use of ProtoDUNE/WCTE (e.g. cosmics rejection in a slow extraction, kinematic reconstruction of final states, etc.) will complement this study.

\section*{Acknowledgments}
This project has received funding from the European Union’s Horizon 2020 Research and Innovation programme under Grant Agreement no. 681647 and the Italian Ministry for Education and Research (MIUR, bando FARE, progetto NUTECH). It is also supported by the Agence Nationale de la Recherche (ANR, France) through the PIMENT project (ANR-21-CE31-0027) and by the Ministry of Science and Education of Republic of Croatia grant No. KK.01.1.1.01.0001

\bibliography{ref.bib}

\begin{thebibliography}{15}%
\makeatletter
\providecommand \@ifxundefined [1]{%
 \@ifx{#1\undefined}
}%
\providecommand \@ifnum [1]{%
 \ifnum #1\expandafter \@firstoftwo
 \else \expandafter \@secondoftwo
 \fi
}%
\providecommand \@ifx [1]{%
 \ifx #1\expandafter \@firstoftwo
 \else \expandafter \@secondoftwo
 \fi
}%
\providecommand \natexlab [1]{#1}%
\providecommand \enquote  [1]{``#1''}%
\providecommand \bibnamefont  [1]{#1}%
\providecommand \bibfnamefont [1]{#1}%
\providecommand \citenamefont [1]{#1}%
\providecommand \href@noop [0]{\@secondoftwo}%
\providecommand \href [0]{\begingroup \@sanitize@url \@href}%
\providecommand \@href[1]{\@@startlink{#1}\@@href}%
\providecommand \@@href[1]{\endgroup#1\@@endlink}%
\providecommand \@sanitize@url [0]{\catcode `\\12\catcode `\$12\catcode `\&12\catcode `\#12\catcode `\^12\catcode `\_12\catcode `\%12\relax}%
\providecommand \@@startlink[1]{}%
\providecommand \@@endlink[0]{}%
\providecommand \url  [0]{\begingroup\@sanitize@url \@url }%
\providecommand \@url [1]{\endgroup\@href {#1}{\urlprefix }}%
\providecommand \urlprefix  [0]{URL }%
\providecommand \Eprint [0]{\href }%
\providecommand \doibase [0]{http://dx.doi.org/}%
\providecommand \selectlanguage [0]{\@gobble}%
\providecommand \bibinfo  [0]{\@secondoftwo}%
\providecommand \bibfield  [0]{\@secondoftwo}%
\providecommand \translation [1]{[#1]}%
\providecommand \BibitemOpen [0]{}%
\providecommand \bibitemStop [0]{}%
\providecommand \bibitemNoStop [0]{.\EOS\space}%
\providecommand \EOS [0]{\spacefactor3000\relax}%
\providecommand \BibitemShut  [1]{\csname bibitem#1\endcsname}%
\let\auto@bib@innerbib\@empty
\bibitem [{\citenamefont {Abi}\ \emph {et~al.}(2020)\citenamefont {Abi} \emph {et~al.}}]{DUNE:2020ypp}%
  \BibitemOpen
  \bibfield  {author} {\bibinfo {author} {\bibfnamefont {B.}~\bibnamefont {Abi}} \emph {et~al.} (\bibinfo {collaboration} {DUNE}),\ }\href@noop {} {\  (\bibinfo {year} {2020})},\ \Eprint {http://arxiv.org/abs/2002.03005} {arXiv:2002.03005 [hep-ex]} \BibitemShut {NoStop}%
\bibitem [{\citenamefont {Abe}\ \emph {et~al.}(2018)\citenamefont {Abe} \emph {et~al.}}]{Hyper-Kamiokande:2018ofw}%
  \BibitemOpen
  \bibfield  {author} {\bibinfo {author} {\bibfnamefont {K.}~\bibnamefont {Abe}} \emph {et~al.} (\bibinfo {collaboration} {Hyper-Kamiokande}),\ }\href@noop {} {\  (\bibinfo {year} {2018})},\ \Eprint {http://arxiv.org/abs/1805.04163} {arXiv:1805.04163 [physics.ins-det]} \BibitemShut {NoStop}%
\bibitem [{\citenamefont {Alekou}\ \emph {et~al.}(2022)\citenamefont {Alekou} \emph {et~al.}}]{Alekou:2022emd}%
  \BibitemOpen
  \bibfield  {author} {\bibinfo {author} {\bibfnamefont {A.}~\bibnamefont {Alekou}} \emph {et~al.},\ }\href {\doibase 10.1140/epjs/s11734-022-00664-w} {\bibfield  {journal} {\bibinfo  {journal} {Eur. Phys. J. ST}\ }\textbf {\bibinfo {volume} {231}},\ \bibinfo {pages} {3779} (\bibinfo {year} {2022})},\ \bibinfo {note} {[Erratum: Eur.Phys.J.ST 232, 15--16 (2023)]},\ \Eprint {http://arxiv.org/abs/2206.01208} {arXiv:2206.01208 [hep-ex]} \BibitemShut {NoStop}%
\bibitem [{\citenamefont {Branca}\ \emph {et~al.}(2021)\citenamefont {Branca}, \citenamefont {Brunetti}, \citenamefont {Longhin}, \citenamefont {Martini}, \citenamefont {Pupilli},\ and\ \citenamefont {Terranova}}]{Branca:2021vis}%
  \BibitemOpen
  \bibfield  {author} {\bibinfo {author} {\bibfnamefont {A.}~\bibnamefont {Branca}}, \bibinfo {author} {\bibfnamefont {G.}~\bibnamefont {Brunetti}}, \bibinfo {author} {\bibfnamefont {A.}~\bibnamefont {Longhin}}, \bibinfo {author} {\bibfnamefont {M.}~\bibnamefont {Martini}}, \bibinfo {author} {\bibfnamefont {F.}~\bibnamefont {Pupilli}}, \ and\ \bibinfo {author} {\bibfnamefont {F.}~\bibnamefont {Terranova}},\ }\href {\doibase 10.3390/sym13091625} {\bibfield  {journal} {\bibinfo  {journal} {Symmetry}\ }\textbf {\bibinfo {volume} {13}},\ \bibinfo {pages} {1625} (\bibinfo {year} {2021})},\ \Eprint {http://arxiv.org/abs/2108.12212} {arXiv:2108.12212 [hep-ex]} \BibitemShut {NoStop}%
\bibitem [{\citenamefont {Longhin}\ \emph {et~al.}(2015)\citenamefont {Longhin}, \citenamefont {Ludovici},\ and\ \citenamefont {Terranova}}]{Longhin:2014yta}%
  \BibitemOpen
  \bibfield  {author} {\bibinfo {author} {\bibfnamefont {A.}~\bibnamefont {Longhin}}, \bibinfo {author} {\bibfnamefont {L.}~\bibnamefont {Ludovici}}, \ and\ \bibinfo {author} {\bibfnamefont {F.}~\bibnamefont {Terranova}},\ }\href {\doibase 10.1140/epjc/s10052-015-3378-9} {\bibfield  {journal} {\bibinfo  {journal} {Eur. Phys. J. C}\ }\textbf {\bibinfo {volume} {75}},\ \bibinfo {pages} {155} (\bibinfo {year} {2015})},\ \Eprint {http://arxiv.org/abs/1412.5987} {arXiv:1412.5987 [hep-ex]} \BibitemShut {NoStop}%
\bibitem [{\citenamefont {Acerbi}\ \emph {et~al.}(2021)\citenamefont {Acerbi} \emph {et~al.}}]{Acerbi:2759849}%
  \BibitemOpen
  \bibfield  {author} {\bibinfo {author} {\bibfnamefont {F.}~\bibnamefont {Acerbi}} \emph {et~al.} (\bibinfo {collaboration} {ENUBET}),\ }\href {https://cds.cern.ch/record/2759849} {\emph {\bibinfo {title} {{NP06/ENUBET Annual Report for the SPSC (2021)}}}},\ \bibinfo {type} {Tech. Rep.}\ (\bibinfo  {institution} {CERN},\ \bibinfo {address} {Geneva},\ \bibinfo {year} {2021})\BibitemShut {NoStop}%
\bibitem [{\citenamefont {Acerbi}\ \emph {et~al.}(2023)\citenamefont {Acerbi} \emph {et~al.}}]{ENUBET:2023hgu}%
  \BibitemOpen
  \bibfield  {author} {\bibinfo {author} {\bibfnamefont {F.}~\bibnamefont {Acerbi}} \emph {et~al.} (\bibinfo {collaboration} {ENUBET}),\ }\href {\doibase 10.1140/epjc/s10052-023-12116-3} {\bibfield  {journal} {\bibinfo  {journal} {Eur. Phys. J. C}\ }\textbf {\bibinfo {volume} {83}},\ \bibinfo {pages} {964} (\bibinfo {year} {2023})},\ \Eprint {http://arxiv.org/abs/2308.09402} {arXiv:2308.09402 [hep-ex]} \BibitemShut {NoStop}%
\bibitem [{\citenamefont {Abi}\ \emph {et~al.}(2017)\citenamefont {Abi} \emph {et~al.}}]{DUNE:2017pqt}%
  \BibitemOpen
  \bibfield  {author} {\bibinfo {author} {\bibfnamefont {B.}~\bibnamefont {Abi}} \emph {et~al.} (\bibinfo {collaboration} {DUNE}),\ }\href@noop {} {\  (\bibinfo {year} {2017})},\ \Eprint {http://arxiv.org/abs/1706.07081} {arXiv:1706.07081 [physics.ins-det]} \BibitemShut {NoStop}%
\bibitem [{\citenamefont {Acerbi}\ \emph {et~al.}(2020)\citenamefont {Acerbi} \emph {et~al.}}]{Acerbi:2020nwd}%
  \BibitemOpen
  \bibfield  {author} {\bibinfo {author} {\bibfnamefont {F.}~\bibnamefont {Acerbi}} \emph {et~al.},\ }\href {\doibase 10.1088/1748-0221/15/08/P08001} {\bibfield  {journal} {\bibinfo  {journal} {JINST}\ }\textbf {\bibinfo {volume} {15}},\ \bibinfo {pages} {P08001} (\bibinfo {year} {2020})},\ \Eprint {http://arxiv.org/abs/2006.07269} {arXiv:2006.07269 [physics.ins-det]} \BibitemShut {NoStop}%
\bibitem [{\citenamefont {Acerbi}\ \emph {et~al.}(2022)\citenamefont {Acerbi} \emph {et~al.}}]{Acerbi:2805716}%
  \BibitemOpen
  \bibfield  {author} {\bibinfo {author} {\bibfnamefont {F.}~\bibnamefont {Acerbi}} \emph {et~al.} (\bibinfo {collaboration} {ENUBET}),\ }\href {https://cds.cern.ch/record/2805716} {\emph {\bibinfo {title} {{NP06/ENUBET annual report 2022 for the SPSC}}}},\ \bibinfo {type} {Tech. Rep.}\ (\bibinfo  {institution} {CERN},\ \bibinfo {address} {Geneva},\ \bibinfo {year} {2022})\ \bibinfo {note} {co-spokesperson F. Terranova}\BibitemShut {NoStop}%
\bibitem [{\citenamefont {Arsenescu}(1999)}]{Arsenescu:1999zs}%
  \BibitemOpen
  \bibfield  {author} {\bibinfo {author} {\bibfnamefont {R.}~\bibnamefont {Arsenescu}} (\bibinfo {collaboration} {SPY/NA56}),\ }\href {\doibase 10.1016/S0920-5632(98)00416-2} {\bibfield  {journal} {\bibinfo  {journal} {Nucl. Phys. B Proc. Suppl.}\ }\textbf {\bibinfo {volume} {70}},\ \bibinfo {pages} {201} (\bibinfo {year} {1999})}\BibitemShut {NoStop}%
\bibitem [{\citenamefont {Atherton}\ \emph {et~al.}(1980)\citenamefont {Atherton}, \citenamefont {Bovet}, \citenamefont {Doble}, \citenamefont {Piemontese}, \citenamefont {Placci}, \citenamefont {Placidi}, \citenamefont {Plane}, \citenamefont {Reinharz}, \citenamefont {Rossa},\ and\ \citenamefont {Von~Holtey}}]{Atherton:133786}%
  \BibitemOpen
  \bibfield  {author} {\bibinfo {author} {\bibfnamefont {H.~W.}\ \bibnamefont {Atherton}}, \bibinfo {author} {\bibfnamefont {C.}~\bibnamefont {Bovet}}, \bibinfo {author} {\bibfnamefont {N.~T.}\ \bibnamefont {Doble}}, \bibinfo {author} {\bibfnamefont {L.}~\bibnamefont {Piemontese}}, \bibinfo {author} {\bibfnamefont {A.}~\bibnamefont {Placci}}, \bibinfo {author} {\bibfnamefont {M.}~\bibnamefont {Placidi}}, \bibinfo {author} {\bibfnamefont {D.~E.}\ \bibnamefont {Plane}}, \bibinfo {author} {\bibfnamefont {M.}~\bibnamefont {Reinharz}}, \bibinfo {author} {\bibfnamefont {E.}~\bibnamefont {Rossa}}, \ and\ \bibinfo {author} {\bibfnamefont {G.}~\bibnamefont {Von~Holtey}},\ }\href {\doibase 10.5170/CERN-1980-007} {\emph {\bibinfo {title} {{Precise measurements of particle production by 400 GeV/c protons on beryllium targets}}}},\ CERN Yellow Reports: Monographs\ (\bibinfo  {publisher} {CERN},\ \bibinfo {address} {Geneva},\ \bibinfo {year} {1980})\BibitemShut {NoStop}%
\bibitem [{\citenamefont {Terranova}\ and\ \citenamefont {Longhin}(2024)}]{Terranova:2896594}%
  \BibitemOpen
  \bibfield  {author} {\bibinfo {author} {\bibfnamefont {F.}~\bibnamefont {Terranova}}\ and\ \bibinfo {author} {\bibfnamefont {A.}~\bibnamefont {Longhin}} (\bibinfo {collaboration} {ENUBET}),\ }\href {https://cds.cern.ch/record/2896594} {\emph {\bibinfo {title} {{NP06/ENUBET annual report 2024 for the SPSC}}}},\ \bibinfo {type} {Tech. Rep.}\ (\bibinfo  {institution} {CERN},\ \bibinfo {address} {Geneva},\ \bibinfo {year} {2024})\BibitemShut {NoStop}%
\bibitem [{\citenamefont {Baratto-Rold\'an}\ \emph {et~al.}(2024)\citenamefont {Baratto-Rold\'an}, \citenamefont {Perrin-Terrin}, \citenamefont {Parozzi}, \citenamefont {Jebramcik},\ and\ \citenamefont {Charitonidis}}]{Baratto-Roldan:2024bxk}%
  \BibitemOpen
  \bibfield  {author} {\bibinfo {author} {\bibfnamefont {A.}~\bibnamefont {Baratto-Rold\'an}}, \bibinfo {author} {\bibfnamefont {M.}~\bibnamefont {Perrin-Terrin}}, \bibinfo {author} {\bibfnamefont {E.~G.}\ \bibnamefont {Parozzi}}, \bibinfo {author} {\bibfnamefont {M.~A.}\ \bibnamefont {Jebramcik}}, \ and\ \bibinfo {author} {\bibfnamefont {N.}~\bibnamefont {Charitonidis}},\ }\href {\doibase 10.1140/epjc/s10052-024-13324-1} {\bibfield  {journal} {\bibinfo  {journal} {Eur. Phys. J. C}\ }\textbf {\bibinfo {volume} {84}},\ \bibinfo {pages} {1024} (\bibinfo {year} {2024})},\ \Eprint {http://arxiv.org/abs/2401.17068} {arXiv:2401.17068 [physics.acc-ph]} \BibitemShut {NoStop}%
\bibitem [{\citenamefont {Garode}\ \emph {et~al.}(2022)\citenamefont {Garode}, \citenamefont {Chinchanikar}, \citenamefont {Garde}, \citenamefont {Mache}, \citenamefont {Joshi}, \citenamefont {Deshmukh}, \citenamefont {Patil}, \citenamefont {Konaka},\ and\ \citenamefont {Hartz}}]{Garode:2022sej}%
  \BibitemOpen
  \bibfield  {author} {\bibinfo {author} {\bibfnamefont {S.~V.}\ \bibnamefont {Garode}}, \bibinfo {author} {\bibfnamefont {S.~S.}\ \bibnamefont {Chinchanikar}}, \bibinfo {author} {\bibfnamefont {C.~S.}\ \bibnamefont {Garde}}, \bibinfo {author} {\bibfnamefont {A.~R.}\ \bibnamefont {Mache}}, \bibinfo {author} {\bibfnamefont {S.~G.}\ \bibnamefont {Joshi}}, \bibinfo {author} {\bibfnamefont {N.}~\bibnamefont {Deshmukh}}, \bibinfo {author} {\bibfnamefont {P.}~\bibnamefont {Patil}}, \bibinfo {author} {\bibfnamefont {A.}~\bibnamefont {Konaka}}, \ and\ \bibinfo {author} {\bibfnamefont {M.}~\bibnamefont {Hartz}} (\bibinfo {collaboration} {WCTE}),\ }\href {\doibase 10.1088/1742-6596/2374/1/012035} {\bibfield  {journal} {\bibinfo  {journal} {J. Phys. Conf. Ser.}\ }\textbf {\bibinfo {volume} {2374}},\ \bibinfo {pages} {012035} (\bibinfo {year} {2022})}\BibitemShut {NoStop}%
\end{thebibliography}%
\end{document}